\documentstyle[11pt]{article}
0

\begin{document}

\begin{center}

~\vfill

{\large \bf  THE AFFINE-METRIC QUANTUM GRAVITY \protect \\
~\vfill
WITH EXTRA LOCAL SYMMETRIES}

\vfill

{\large M. Yu. Kalmykov}
\footnote { E-mail: $kalmykov@thsun1.jinr.dubna.su$}

\vspace{1cm}

{\em Bogoliubov Laboratory of Theoretical Physics,
 Joint Institute for Nuclear  Research,
 $141~980$ Dubna $($Moscow Region$)$, Russian Federation} \protect \\

\vspace{1cm}

\end{center}

\vfill

\begin{abstract}
We discuss the role of additional local symmetries related to the
transformations of connection fields in the affine-metric theory
of gravity. The corresponding BRST transformations connected with
all symmetries (general coordinate, local Lorentz and extra) are
constructed.  It is shown, that extra symmetries give the additional
contribution to effective action which is proportional to the
corresponding Nielsen-Kallosh ghost one. Some arguments are given,
that there is no anomaly associated with extra local symmetries.
\end{abstract}

\vfill

PACS number(s) 0450+h, 0460-m

\pagebreak

\section{Introduction}

At present, there is no theory of gravity that would be satisfactory
>from the viewpoint of quantum field theory. The Einstein gravity,
although agrees with all the available experimental data at the
classical level, is not renormalizable. For the pure Einstein gravity,
the one-loop counterterms vanish on shell \cite{THV} but the two-loop
counterterms break the renormalizability of the model \cite{GS}.
Furthermore, as one adds matter fields, the renormalizability is
violated already at the one-loop level \cite{DPN1}.

Therefore, one has either to modify the theory or to prove that the
difficulties are due to imperfection of the methods applied to treat
the model.

We are inclined to believe perturbative renormalizability to be one of the
fundamental criteria for a "true" quantum gravity. We are going to assume
that a perturbative renormalizable gravity exists but may differ from the
Einstein theory.

The simplest way to modifying the Einstein gravity consists in adding some
terms to the action which are quadratic in the curvature. This kind of a
theory would be multiplicatively renormalizable and asymptotically free but
not
unitary \cite{St2}. Ghosts and tachyons would be present in the spectrum of
the tree-level $S-$matrix, owing to the $\frac{1}{p^4}$ behavior of the
metric field propagator, because the action involves higher derivatives.
Attempts to restore unitarity by taking into account quantum corrections
or adding some matter fields, failed \cite{Ant}.

We believe that the situation may be saved by increasing the symmetries of
the initial action. Additional global or local symmetries that are maintained
at the quantum level without generating anomalies may essentially improve
renormalization properties. An example is supersymmetry. The simplest $N = 1$
supergravity is the first theory of gravity with matter in which the one-loop
$S-$matrix is ultraviolet finite \cite{SUGRA1}. The known models of the
$N = 1$ supergravity are finite up to two loops but may generate nonvanishing
three-loop divergent counterterms. Models with the extended $(e.g. N = 8)$
supersymmetry or some other additional symmetry (e.g. the local conformal
symmetry)  have better renormalization features \cite{SUGRA2}, but there is
no proof of their complete finiteness by now.

Another possibility is that space-time has a reacher structure than just a
Riemann space. This implies introducing, besides the metric
$g_{\mu \nu }$, some other geometric structures like torsion and nonmetricity.

The so-called Riemann-Cartan space is the simplest extension. It has
the torsion as an independent dynamical variable. There is a wide literature
on the gravity with torsion \cite{FWH1}. We only remind some facts
concerning the
renormalization properties. In the framework of the Riemann-Cartan space, one
succeeded in constructing the models that are unitary at the tree level
in the linear approximation \cite{Nev1}. Independent dynamical variables,
metrics and torsion, have propagators with the $\frac{1}{p^2}$ behavior.
However, at the one-loop level, the dimension-$4$ counterterms should be
generated which are not forbidden by the symmetries present in the model,
i.e. any scalars constructed from the contractions of the curvature and
torsion tensors. In general, these counterterms will break either
renormalizability or unitarity.

A further extension involves the affine-metric space-time characterized by
the metric $g_{\mu \nu}$ and the affine connection
$\bar{\Gamma}^\sigma_{~\mu \nu}$ \cite{Hehl1a}.
There are several variants proposed by now \cite{HKH-76}.
In the framework of the affine-metric theory there exist about two hundred
arbitrary coefficients , which are not defined from basic principles. Because
of the technical complication of this model, the renormalizability properties
of the affine-metric quantum gravity have been studied insufficiently.

New hopes for a more perfect quantum gravity arose in connection with
strings. The discussion of the bosonic string on the affine-metric
manifold is given in \cite{Tarasov}.

Summarize the basic principles of the "true" quantum gravity:
\begin{itemize}
\item
Basic assumption is the existence of a perturbatively renormalizable
theory of gravity. As a consequence, all the assertions of the conventional
perturbative quantum field theory should remain valid. In particular, the
theory should be unitary at the tree level. Since no purely metric gravity
is known to be both renormalizable and unitary, we are bound to introduce
additional dynamical variables or to impose some new local symmetry, like
supersymmetry. Thus, the true quantum perturbatively renormalizable, unitary
theory of gravity without the supersymmetry can only be constructed in the
framework of the extended space-time geometry, like the Riemann-Cartan or
affine-metric geometry.
\item
The Lagrangian should not involve higher derivatives of the fields, in order
to avoid ghosts and tachyons in the spectrum.
\item
The renormalization group analysis shows \cite{ren} that in an ordinary
renormalizable quantum field theory the most essential role belongs to the
terms with the dimension of the space-time. Consequently, it is natural to
start constructing a renormalizable model from the terms with the
corresponding dimension.
\item
For the classical limit, coinciding with the Einstein theory, to exist
one needs to add a term linear in the curvature tensor to the Lagrangian.
\end{itemize}

In this paper, we consider the affine-metric gravity. This theory may be
invariant with respect to some extra local transformations of the affine
connection for special value of the coefficients \cite{Bars}-\cite{MKL}.
These invariances restrict arbitrariness of the initial Lagrangian and avoid
undesirable counterterms at the quantum level. However, the quantum
properties of these symmetries have not been studied.

The main aim of this paper is to investigate extra local symmetries
connected with the local transformation of the affine-metric gravity at
the quantum level. We would like to show that  these symmetries may
give additional contribution to the effective action.  The
BRST-transformations connected with the symmetries of the affine-metric
theory are constructed as a basis for further investigations of
renormalizability and unitary properties. We don't discuss the physical
ground and geometric nature of these symmetries in this paper.  In
section $2$, we give the structure of the affine-metric gravity and
introduce extra symmetries. In section $3$, we introduce the
BRST-transformations connected with extra symmetries in the geometrical
and tetrad approaches, construct the quantum Lagrangian and discuss the
problem of anomalies connected with new symmetries. In section $4$, we
conclude with a discussion of the results and perspective.

The following notation and conventions are accepted:

$$ c = \hbar = 1;~~~~~ \mu , \nu  = 0,1,2,3;~~~~~ {\it k}^2 = 16 \pi
G, ~~~~~\varepsilon  = \frac{4-d}{2} $$
$$~~~~~(g) = det(g_{\mu \nu }),~~~~~ e = det(e^a_{~\mu})
~~~~~\eta_{\mu \nu } = (+ - - -),$$

The Riemannian connection is
$\Gamma^\sigma_{~\mu \nu } = g^{\sigma \lambda} \frac{1}{2}
\left(- \partial_\lambda g_{\mu \nu} + \partial_\mu g_{\lambda \nu}
+ \partial_\nu g_{\lambda \mu} \right)$.~
Objects marked by bar are constructed by means of the affine connection
$\bar \Gamma^\sigma_{~\mu \nu}$. The others are the Riemannian objects.
For further calculations one needs to define the following tensor object:
$ D^\sigma_{~\mu \nu }  = \bar{\Gamma}^\sigma_{~\mu \nu }
- \Gamma^\sigma_{~\mu \nu }$.

\section{Extra local symmetries in affine-metric gravity}

The affine-metric manifold permits the geometric and tetrad description.
The geometric approach implies the description in terms of the metric
$g_{\mu \nu}$ and affine connection $\bar{\Gamma}^\sigma_{~\mu \nu}$.
The basic objects are expressed as

\begin{itemize}
\item
curvature
\begin{equation}
\bar{R}^\sigma_{~\lambda \mu \nu } (\bar{\Gamma}) =
\partial_\mu \bar{\Gamma}^\sigma_{~\lambda \nu}
 -
\partial_\nu \bar{\Gamma}^\sigma_{~\lambda \mu}
+ \bar{\Gamma}^\sigma_{~\alpha \mu} \bar{\Gamma}^\alpha_{~\lambda \nu}
- \bar{\Gamma}^\sigma_{~\alpha \nu} \bar{\Gamma}^\alpha_{~\lambda \mu}
\label{acur}
\end{equation}
\item
torsion
\begin{equation}
\bar{Q}^\sigma_{~\mu \nu }(\bar{\Gamma}) = \frac{1}{2}
\left( \bar \Gamma^\sigma_{~ \mu \nu} - \bar{\Gamma}^\sigma_{~\nu \mu }
\right)
\label{ator}
\end{equation}
\item
nonmetricity
\begin{equation}
\bar{W}_{\sigma \mu \nu}(g, \bar{\Gamma})
= \bar{\nabla}_\sigma g_{\mu \nu } = \partial_\sigma g_{\mu \nu }
- \bar{\Gamma}^\alpha_{~\mu \sigma} g_{\alpha \nu}
- \bar{\Gamma}^\alpha_{~\nu \sigma} g_{\alpha \mu}
\label{anon}
\end{equation}
\end{itemize}

In the tetrad formalism, for describing the manifold we use the tetrad
$e^a_{~\mu}$  and the local Lorentz connection $\bar{\Omega}^a_{~b \mu}$.
Using the following relations \cite{Hehl1a}

\begin{equation}
g_{\mu \nu} = e^a_{~\mu} e^b_{~\nu} \eta_{a b}
\label{connection1}
\end{equation}

\begin{equation}
\bar{\nabla}_\sigma e^a_{~\mu} = \partial_\sigma e^a_{~\mu} +
\bar{\Omega}^a_{~b \sigma} e^b_{~\mu} - \bar{\Gamma}^\nu_{~\mu \sigma}
e^a_{~\nu} = 0
\label{connection2}
\end{equation}
where $\eta_{a b}$ is the Minkowskian metric,
we can obtain the main geometric objects in tetrad formalism:

\begin{itemize}
\item
curvature
\begin{equation}
\bar{R}^\sigma_{~\lambda \mu \nu } (\bar{\Gamma}) =
\bar{R}^a_{~b \mu \nu } (\bar{\Omega})  e_a^{~\sigma} e^b_{~\lambda} =
( \partial_\mu \bar{\Omega}^a_{~b \nu}
 -
\partial_\nu \bar{\Omega}^a_{~b \mu}
+ \bar{\Omega}^a_{~\alpha \mu} \bar{\Omega}^\alpha_{~b \nu}
- \bar{\Omega}^a_{~\alpha \nu} \bar{\Omega}^\alpha_{~b \mu} )
e_a^{~\sigma} e^b_{~\lambda}
\end{equation}

\item
torsion
\begin{equation}
\bar{Q}^\sigma_{~\mu \nu }(\bar{\Gamma}) =
\bar{Q}^a_{~\mu \nu }(e,\bar{\Omega}) e_a^{~\sigma} =
-\frac{1}{2}
\left(
\partial_\mu e^a_{~\nu}  - \partial_\nu e^a_{~\mu}  +
\bar{\Omega}^a_{~b \mu} e^b_{~\nu}  -
\bar{\Omega}^a_{~b \nu} e^b_{~\mu} \right) e_a^{~\sigma}
\end{equation}
\item
nonmetricity
\begin{equation}
\bar{W}_{\sigma \mu \nu}(g, \bar{\Gamma}) =
\bar{W}_{\sigma a b }(\bar{\Omega}) e^a_{\mu} e^b_{\nu}
= - \left( \bar{\Omega}_{ab \sigma} + \bar{\Omega}_{ba \sigma} \right)
e^a_{\mu} e^b_{\nu}
\end{equation}
\end{itemize}

An affine-metric theory of gravity may have additional local symmetries
related to transformations of the connection \cite{Bars}-\cite{MKL}.
The simplest is the transformation of irreducible parts of the
connection. The affine connection can be rewritten as

\begin{equation}
\bar{\Gamma}^\sigma_{~\mu \nu} =
\Gamma^\sigma_{~\mu \nu} + D^\sigma_{~\mu \nu}
\end{equation}
where $D^\sigma_{~\mu \nu }$ is the tensor.
An arbitrary tensor of third rank  $D_{\sigma \mu \nu }$
is known to be expandable in terms of the following irreducible parts:

\begin{equation}
D_{\sigma \mu \nu }  = A_\sigma g_{\mu \nu }  + B_\mu g_{\nu \sigma }
+ C_\nu g_{\mu \sigma }  + \frac{1}{6}
\check{D}_{[\sigma \mu \nu]}  +
\underline{D}_{\sigma \mu \nu }
\end{equation}
\noindent where $\check{D}_{[\sigma \mu \nu]}$ is the antisymmetric part;
$A_\sigma ,B_\mu$ and $C_\nu ,$ are the vector fields

\begin{equation}
A_\sigma  \equiv \frac{1}{18} \left( 5D_{\sigma \lambda }^{~~\lambda }
- D^\lambda_{~\sigma \lambda } - D^\lambda_{~\lambda \sigma }  \right)
\end{equation}

\begin{equation}
B_\sigma  \equiv \frac{1}{18} \left( -  D_{\sigma \lambda }^{~~\lambda
 } + 5 D^\lambda_{~\sigma \lambda } - D^\lambda_{~\lambda \sigma }
\right)
\end{equation}

\begin{equation}
C_\sigma  \equiv \frac{1}{18} \left( - D_{\sigma \lambda }^{~~\lambda }
- D^\lambda_{~\sigma \lambda } + 5 D^\lambda_{~\lambda \sigma }
\right)
\end{equation}
\noindent
and $\underline{D}_{\sigma \mu \nu }$ is the traceless part satisfying the
following conditions:

\begin{equation}
\underline{D}^\nu_{~\mu \nu }  = \underline{D}^\nu_{~\nu \mu }
= \underline{D}_{\nu \mu }^{~~\mu } \equiv 0
\end{equation}

\begin{equation}
\epsilon^{\lambda \sigma \mu \nu} \underline{D}_{\sigma \mu \nu } = 0
\end{equation}

The symmetries related to transformations of irreducible parts are

\begin{equation}
\bar{\Gamma}^\sigma_{~ \mu \nu } \rightarrow '\bar{\Gamma}^\sigma_{~ \mu \nu}
= \bar{\Gamma}^\sigma_{~ \mu \nu } + g^{\sigma \rho} \Lambda_{[\rho \mu \nu]}
+ g^{\sigma \rho} \underline{T}_{\rho \mu \nu } + M^\sigma g_{\mu \nu }  +
N_\nu \delta^\sigma_\mu  + P_\mu g^\sigma_\nu
\label{testtr}
\end{equation}
where $\Lambda_{[\sigma \mu \nu] },
\underline{T}_{\sigma \mu \nu },M_\sigma, N_\nu, P_\mu
$ are arbitrary antisymmetric, traceless tensors and vectors, respectively.
If the theory is invariant under all the
symmetries (\ref{testtr}), the affine connection has no dynamical
degrees  of freedom. Just becomes an auxiliary field. After eliminating it
the theory reduced to a metric gravity.
We restrict ourselves to considering only some particular cases of
(\ref{testtr}):

\begin{equation}
\bar{\Gamma}^\sigma_{~\mu \nu} \rightarrow
'\bar{\Gamma}^\sigma_{~\mu \nu} =
\bar{\Gamma}^\sigma_{~ \mu \nu} + g^{\sigma \lambda}
\Lambda_{[\lambda \mu \nu ]}(x) + \delta^\sigma_\mu C_\nu(x)
\label{comb}
\end{equation}
This transformation is the sum of the projective transformations
\cite{project}:

\begin{equation}
\bar{\Gamma}^\sigma_{~\mu \nu} \rightarrow
'\bar{\Gamma}^\sigma_{~\mu \nu} =
\bar{\Gamma}^\sigma_{~ \mu \nu} + \delta^\sigma_\mu C_\nu(x)
\label{projec}
\end{equation}
and antisymmetric transformations:

\begin{equation}
\bar{\Gamma}^\sigma_{~\mu \nu} \rightarrow
'\bar{\Gamma}^\sigma_{~\mu \nu} =
\bar{\Gamma}^\sigma_{~ \mu \nu} + g^{\sigma \lambda}
\Lambda_{[\lambda \mu \nu ]}(x)
\label{ant}
\end{equation}

Consider the tetrad approach. Using relations (\ref{connection2}) it is easy
to show that the transformation (\ref{comb}) has the following form in the
tetrad formalism:

\begin{equation}
\bar{\Omega}^a_{~b \sigma} \rightarrow '\bar{\Omega}^a_{~b \sigma} =
\bar{\Omega}^a_{~b \sigma} + e^{a \lambda}
\Lambda_{[\lambda b \sigma]}(x) + \delta^a_b C_\sigma(x)
\label{comb1}
\end{equation}

What is the meaning of these extra gauge invariances? Usually, local
invariances are related to physically relevant groups like, for
instance, the diffeomorphism and Lorentz group or internal group in the
Yang-Mills type theories. And usually, a gauge field (potential,
connection) is adjoined to each gauge invariance. However, for extra
local symmetries no additional gauge fields appear. Therefore, these
symmetries do not fit into the framework of "ordinary" gauge theories.
This is a gauge invariance without of any physical meaning and
geometric nature.  At the same time, the presence of extra gauge
invariances imposes some new constraints on the source terms.
These new constraints may solve some old problems concerning the
interaction of matter and vector gauge fields with connection in
affine-metric gravity \cite{Hehl1a}, \cite{HKH-76}.
In our opinion, the role of these symmetries infers in
excluding counterterms of the particular type and restricting
arbitrariness of the initial Lagrangian. This is possible if the
symmetries are maintained at the quantum level. The questions arise
about the gauge fixing and corresponding ghost fields connected with
these symmetries. For constructing the quantum Lagrangian we must add
the gauge fixing terms and appropriate Faddeev-Popov ghost fields.
We derive the corresponding theory from the invariance of the full
Lagrangian under the BRST-transformations

\begin{equation}
 {\bf s} {L}_{quan}  = 0
\end{equation}
where {\bf s} is a graded, nilpotent BRST operator.

\section{BRST-transformations in the affine-metric gravity with extra local
invariances}

Consider an arbitrary model of the affine-metric gravity quadratic in
the torsion, curvature and nonmetricity, invariant under the general
coordinate and additional local transformations (\ref{comb}) in the
geometric approach. The propagators which stem from the classical
Lagrangian are not all defined because the quadratic field
approximation of the initial Lagrangian is degenerated, i.e. contains
modes associated with the gauge invariance. The propagators can be made
invertible if a gauge fixing term is added to initial Lagrangian. We
consider the symmetries (\ref{comb}) as a gauge symmetry. Hence, we
must fix these symmetries. The gauge fixing Lagrangian is

\begin{equation}
L_{gf}  = \left( b^\mu \omega_{\mu \nu} F^\nu
+ \pi^\mu \zeta_{\mu \nu } f^\nu
+ d^\mu \varsigma _{\mu \nu }\rho^\nu
- \frac{1}{2} b^\mu \omega_{\mu \nu } b^\nu
- \frac{1}{2} \pi^\mu \zeta_{\mu \nu } \pi^\nu
- \frac{1}{2} d^\mu \varsigma_{\mu \nu } d^\nu \right) \sqrt{-g}
\label{fixing}
\end{equation}
where $\{ b_\mu$, $\pi_\mu$, $d_\mu \}$ are auxiliary fields,
$\{ \omega_{\mu \nu }, \varsigma_{\mu \nu }, \zeta_{\mu \nu } \}$
are arbitrary differential operators which contain two or smaller
derivatives and
$\{ F_\mu$, $f_\mu$, $\rho_\mu \}$ fix the general coordinate, projective
and antisymmetric gauges.

Since auxiliary fields appear without derivatives in the Lagrangian,
they can be eliminated by means of their equations of motion which
yield

\begin{equation}
L_{gf}  = \left( \frac{1}{2} F^\mu \omega_{\mu \nu } F^\nu
+ \frac{1}{2} f^\mu \zeta_{\mu \nu } f^\nu
+ \frac{1}{2} \rho^\mu \varsigma_{\mu \nu } \rho^\nu \right)
\sqrt{-g}
\end{equation}

We fix the general coordinate transformations by the following gauge:

\begin{equation}
F^\mu = (-g)^{-\tau} \partial_\nu \left( g^{\mu \nu} (-g)^\tau \right)
+ a_1 \bar{\Gamma}^\mu_{~\alpha \beta} g^{\alpha \beta}
+ a_2 \bar{\Gamma}^\nu_{~\alpha \nu} g^{\alpha \mu}
+ a_3 \bar{\Gamma}^\nu_{~\nu \alpha} g^{\alpha \mu}
\label{coordinate}
\end{equation}
where $\tau$ and $a_1,a_2, a_3$ are arbitrary constants.
The  projective gauge (\ref{projec}) can be fixed as follows \cite{MKL}:

\begin{equation}
f^\lambda = \left(f_1 \delta_\sigma^\lambda g^{\mu \nu}
+  f_2 g^{\mu \lambda} \delta^\nu_\sigma
+ f_3 g^{\nu \lambda} \delta^\mu_\sigma \right)
\bar{\Gamma}^\sigma_{~\mu \nu}
\label{general}
\end{equation}
where $\{f_i\}$ are constants satisfying the condition:

\begin{equation}
f_1 + f_2 + 4f_3 \neq 0
\label{aux}
\end{equation}

The most general coordinate and projective gauge fixing terms are

\begin{equation}
F^\mu = (-g)^{-\tau} \partial_\nu \left( g^{\mu \nu} (-g)^\tau \right)
+ a_1 \bar{\Gamma}^\mu_{~\alpha \beta} g^{\alpha \beta}
+ a_2 \bar{\Gamma}^\nu_{~\alpha \nu} g^{\alpha \mu}
+ a_3 \bar{\Gamma}^\nu_{~\nu \alpha} g^{\alpha \mu}
+ E^{\mu \rho \lambda ~ \alpha \beta}_{~~~~\sigma} \nabla_\rho \nabla_\lambda
\bar{\Gamma}^\sigma_{~\alpha \beta}
\label{coordinate1}
\end{equation}

\begin{equation}
f^\lambda = \left(f_1 \delta_\sigma^\lambda g^{\mu \nu}
+  f_2 g^{\mu \lambda} \delta^\nu_\sigma
+ f_3 g^{\nu \lambda} \delta^\mu_\sigma \right)
\bar{\Gamma}^\sigma_{~\mu \nu}
+ G^{\lambda \rho \lambda ~ \alpha \beta}_{~~~~ \sigma}
\nabla_\rho \nabla_\lambda \bar{\Gamma}^\sigma_{~\alpha \beta}
\label{general1}
\end{equation}
where $E^{\mu \rho \lambda ~ \alpha \beta}_{~~~~\sigma} $ and
$G^{\mu \rho \lambda ~ \alpha \beta}_{~~~~\sigma} $ are reduction tensors,
i.e. a product of metric tensors and Kroneker's symbols. In the initial
Lagrangian the independent dynamical fields have propagators with the
$\frac{1}{p^2}$ behavior. Then, the gauge-fixing terms proportional to
$E^{\mu \rho \lambda ~ \alpha \beta}_{~~~~\sigma}$ and
$G^{\mu \rho \lambda ~ \alpha \beta}_{~~~~\sigma}$ with higher derivatives
may break the unitarity of the theory. To avoid this problem we consider the
case $E^{\mu \rho \lambda ~ \alpha \beta}_{~~~~\sigma} =
G^{\mu \rho \lambda ~ \alpha \beta}_{~~~~\sigma} = 0$.

For the antisymmetric transformation (\ref{ant})  we use the gauge condition

\begin{equation}
\rho^\lambda = \epsilon^{\lambda \sigma \mu \nu} \bar{\Gamma}_{\sigma \mu \nu}
\label{axial}
\end{equation}

The BRST-transformations are obtained in the usual way \cite{baulieu}
>from gauge transformations by replacing the gauge parameter by the
corresponding ghost field

\begin{eqnarray}
{\bf s} g_{\mu \nu} & = & {\cal L}_c g_{\mu \nu}
~~~~~{\bf s} \bar{\Gamma}^\sigma_{~\mu \nu}  =
{\cal L}_c \bar{\Gamma}^\sigma_{~\mu \nu}
+ {\it k} \delta^\sigma_\mu \chi_\nu
+ {\it k} \eta^\sigma_{~\mu \nu }
\nonumber \\
{\bf s} \bar{c}_\mu  & = & b_\mu
~~~~~~~~~~{\bf s} b_\mu  =  0
~~~~~~~{\bf s} c^\mu   =  c^\lambda \partial_\lambda c^\mu
\nonumber \\
{\bf s} \bar{\chi}_\mu & = & \pi_\mu
~~~~~~~~~~{\bf s} \pi_\mu = 0
~~~~~~{\bf s} \chi^\nu = 0
\nonumber \\
{\bf s} \bar{\eta}_\mu & = & d_\mu
~~~~~~~~~~{\bf s} d_\mu = 0
~~~~~~{\bf s} \eta^\sigma_{~\mu \nu } = 0
\label{full-BRST}
\end{eqnarray}
where ${\bf s}$ is a graded, nilpotent BRST operator and
$\{\bar{c}_\nu, c^\mu,\}$, $\{ \bar{\chi}_\alpha, \chi^\beta \},
\{\bar{\eta}_\sigma, \eta^\lambda_{~\mu \nu } \}$
are anticommuting ghost fields connected with general coordinate,
projective and antisymmetric transformations, respectively;
${\cal L}_\xi A^{\mu_1 \ldots \mu_k}_{~~~~~\nu_1 \ldots \nu_l}$
is an ordinary Lie derivative. Under the general coordinate transformations
$x^\mu \rightarrow 'x^\mu = x^\mu + {\it k} \xi^\mu$,the Lie derivatives are:

\begin{eqnarray}
{\cal L}_\xi (-g)^\alpha  & = &
 {\it k} \alpha
\left(2 \partial_\sigma \xi^\sigma
+ \xi^\sigma g^{\mu \nu} \partial_\sigma g_{\mu \nu}\right) (-g)^\alpha
+ O ({\it k}^2)
\nonumber \\
{\cal L}_\xi g_{\mu \nu}  & = &
 {\it k} \left(\partial_\mu \xi^\beta g_{\beta \nu}
+ \partial_\nu \xi^\beta g_{\beta \mu}
+ \xi^\beta \partial_\beta g_{\mu \nu}\right)
+ O ({\it k}^2)
\nonumber \\
{\cal L}_\xi \bar{\Gamma}^\sigma_{~\mu \nu}(x) & = &
{\it k} \left(
\bar{\Gamma}^\sigma_{~\alpha \nu} \partial_\mu \xi^\alpha
+
\bar{\Gamma}^\sigma_{~\mu \alpha } \partial_\nu \xi^\alpha
-
\bar{\Gamma}^\alpha_{~\mu \nu} \partial_\alpha \xi^\sigma
+ \xi^\alpha \partial_\alpha \bar{\Gamma}^\sigma_{~\mu \nu}
+ \partial_{\mu \nu} \xi^\sigma \right)
\nonumber \\
& + & O ({\it k}^2)
\end{eqnarray}

The action of ${\bf s}$ on any function of fields is given by the graded
Leibniz rule:

\begin{eqnarray}
{\bf s} \left( XY \right) & = & \left( {\bf s}X \right) Y \pm
X \left( {\bf s}Y \right)
\nonumber \\
{\bf s} \partial_\mu & = & \partial_\mu {\bf s}
\nonumber \\
{\bf s}^2 & = & 0
\end{eqnarray}
where the minus sign occurs if X contains an odd number of ghosts and
anti-ghosts.

The quantum Lagrangian is
\begin{equation}
L_{quant} = L_{clas} + {\bf s} \biggl( \bar{c}^\mu \omega_{\mu \nu}
\left( F^\nu - \frac{1}{2} b^\nu \right) + \bar{\chi}^\mu \zeta_{\mu \nu}
\left( f^\nu - \frac{1}{2} \pi^\nu \right)
+ \bar{\eta}^\mu \varsigma_{\mu \nu}
\left( \rho^\nu - \frac{1}{2} d^\nu \right) \biggr)
\end{equation}
where for simplicity we consider the case ${\bf s}\omega_{\mu \nu } =
{\bf s}\zeta_{\mu \nu } = {\bf s}\varsigma_{\mu \nu } = 0$. The generating
functional is

\begin{equation}
e^{i W} =
\int dg_{\mu \nu }~ d\bar{\Gamma}^\sigma_{~\mu \nu }
~d \bar{c}^\mu ~dc^\nu ~d \bar{\chi}^\mu ~d\chi^\nu
~d \bar{\eta}^\mu ~d\eta^{\sigma \lambda \nu}
e^{i S_{quan}}
\left(det \omega_{\mu \nu } \right)^{\frac{1}{2}}
\left(det \zeta_{\mu \nu } \right)^{\frac{1}{2}}
\left(det \varsigma_{\mu \nu } \right)^{\frac{1}{2}}
\end{equation}
where
$det \omega_{\mu \nu }, det \zeta_{\mu \nu },  det \varsigma_{\mu \nu }$
are the so called Nielsen-Kallosh ghosts.

>From (\ref{fixing}) and (\ref{full-BRST}) we obtain the one-loop ghost
Lagrangian:

\begin{eqnarray}
 L_{gh}  & = &   - \bar{c}^\mu \omega_{\mu \nu} {\bf s} F^\nu
- \bar{\chi}^\mu \zeta_{\mu \nu } {\bf s} f^\nu
- \bar{\eta}^\mu \varsigma_{\mu \nu } {\bf s} \rho^\nu
\nonumber \\
& = &
- \left( \bar{c}^\mu ~\bar{\chi}^\mu ~\bar{\eta}^\mu \right)
\left( \begin{array}{ccc}
\omega_{\mu \sigma} \triangle^\sigma_{~\nu}
& \omega_{\mu \nu} (a_1 + a_2 + 4a_3)& 0 \\
\zeta_{\mu \sigma} Z^\sigma_{~\nu} &  (f_1 + f_2 + 4f_3) \zeta_{\mu \nu}
& 0 \\
\varsigma_{\mu \sigma} L^\sigma_{~\nu} & 0
& \varsigma_\mu^{~ \alpha} \epsilon_{\alpha \sigma \lambda \nu}
\end{array} \right)
\left( \begin{array}{c}
c^\nu \\
\chi^\nu \\
\eta^{\sigma \lambda \nu }
\end{array} \right)
\label{gh}
\end{eqnarray}
where  $\triangle^\mu_{~\nu}, Z^\mu_{~ \nu}$ and $L^\mu_{~\nu}$ are

\begin{eqnarray}
\triangle_{\mu \nu} & = &
{\it k}  \biggl(
\left(a_1 -1 \right) g_{\mu \nu} \nabla ^2
+ \frac{1}{2} \left( a_2 + a_3 + 2 \tau - 1 \right)
\left( \nabla_\mu \nabla_\nu + \nabla_\nu \nabla_\mu \right)
\nonumber \\
&&
+ \frac{1}{2} \left( 2a_1 - 1 - a_2 - a_3 - 2 \tau \right) R_{\mu \nu}
+  \left( a_1 \nabla_\nu
D^{~~\lambda}_{\mu \lambda} + a_2 \nabla_\nu D^\sigma_{~\mu \sigma} +
a_3 \nabla_\nu D^\lambda_{~\lambda \mu} \right)
\nonumber \\
&&
+ \left( a_2 D^\lambda_{~\nu \lambda}
+ a_3 D^\lambda_{~\lambda \nu} \right) \nabla_\mu
-  a_1  \left( g_{\mu \nu} D^{\alpha \beta}_{~~~\beta}
  \nabla_\alpha
  - D_{\mu \nu}^{~~~\lambda} \nabla_\lambda
- D_{\mu ~ \nu}^{~ \lambda} \nabla_\lambda  \right) \biggr)
\nonumber \\
&&
+ O({\it k}^2)
\end{eqnarray}

\begin{eqnarray}
Z_{\mu \nu} & = & {\it k} \biggl(
f_1 g_{\mu \nu} \nabla ^2  + \frac{1}{2} \left( f_2 + f_3 \right)
\left( \nabla_\mu \nabla_\nu + \nabla_\nu \nabla_\mu \right)
+ \frac{1}{2} \left( 2f_1 - f_2 - f_3 \right) R_{\mu \nu}
\nonumber \\
& + &   \left( f_1 \nabla_\nu
D^{~~\lambda}_{\mu \lambda} + f_2 \nabla_\nu D^\sigma_{~\mu \sigma} +
f_3 \nabla_\nu D^\lambda_{~\lambda \mu} \right)
+ \left( f_2 D^\lambda_{~\nu \lambda}
+ f_3 D^\lambda_{~\lambda \nu} \right) \nabla_\mu
\nonumber \\
& - & f_1  \left( g_{\mu \nu} D^{\alpha \beta}_{~~~\beta}
  \nabla_\alpha
  - D_{\mu \nu}^{~~~\lambda} \nabla_\lambda
- D_{\mu ~ \nu}^{~ \lambda} \nabla_\lambda  \right) \biggr)
+ O({\it k}^2)
\end{eqnarray}

\begin{equation}
L^\mu_{~ \nu}  =  {\it k} \epsilon^{\mu \alpha \beta \lambda}
\left( \bar{Q}_{\alpha \beta \nu} \nabla_\lambda
+ D_{\alpha \nu \beta} \nabla_\lambda
- g_{\alpha \nu} D^\sigma_{~\beta \lambda} \nabla_\sigma
+ \nabla_\nu Q_{\alpha \beta \lambda} \right) + O({\it k}^2)
\end{equation}

Let us consider the case

\begin{equation}
a_1 + a_2 + 4a_3 = 0
\end{equation}
Then, we can get the diagonal form of the ghost Lagrangian (\ref{gh})
by the following redefinition of the ghost fields:

\begin{eqnarray}
\tilde{\chi}^\nu & = & \chi^\nu + \frac{1}{f_1 + f_2 + 4f_3}
Z^\nu_{~\sigma} c^\sigma
\nonumber \\
\tilde{\eta}^{\sigma \mu \nu } & =  &
\eta^{\sigma \mu \nu } + \frac{1}{6} \epsilon^{\sigma \mu \nu \lambda }
Z_{\lambda \alpha } c^\alpha
\end{eqnarray}

This redefinition does not change the functional integral measure. In the
new variables the ghost Lagrangian has the diagonal form:

\begin{equation}
L_{gh} =
- \left( \bar{c}^\mu ~\bar{\chi}^\mu ~\bar{\eta}^\mu \right)
\left( \begin{array}{ccc}
\omega_{\mu \sigma} \triangle^\sigma_{~\nu} & 0 & 0 \\
0 &  (f_1 + f_2 + 4f_3) \zeta_{\mu \nu} & 0 \\
0 & 0 & \varsigma_\mu^{~ \alpha} \epsilon_{\alpha \sigma \lambda \nu}
\end{array} \right)
\left( \begin{array}{c}
c^\nu \\
\tilde{\chi}^\nu \\
\tilde{\eta}^{\sigma \lambda \nu }
\end{array} \right)
\end{equation}

The loop contribution of the projective and antisymmetric ghosts to the
effective action is proportional to $\left( det \zeta_{\mu \nu } \right)
\left( det \varsigma_{\alpha \beta} \right)$. The one-loop generating
functional is

\begin{equation}
e^{i W} =
\int dg_{\mu \nu }~ d\bar{\Gamma}^\sigma_{~\mu \nu }
e^{i \left( S_{clas} + S_{gf} \right) }
\left(det \omega_{\mu \nu } \right)^\frac{1}{2}
\left(det \omega_{\mu \alpha} \triangle^\alpha_{~\nu}  \right)
\left(det \zeta_{\mu \nu } \right)^\frac{3}{2}
\left(det \varsigma_{\mu \nu } \right)^\frac{3}{2}
\label{result1}
\end{equation}

In this way, in the geometric formalism the projective and
antisymmetric ghost contribution is added to the corresponding
Nielsen-Kallosh ghost one.  Hence, the presence of extra symmetries,
which have not the physical meaning, give a new, extra contribution to
the effective action. This contribution may improve the renormalizable
properties of the theory.

In an analogous way consider the affine-metric gravity in the tetrad
formalism.  In the tetrad formalism the theory is invariant under the
general coordinate, local Lorentz and additional (\ref{comb1})
transformations. Let us construct the corresponding BRST-symmetry. The
gauge fixing Lagrangian looks like

\begin{eqnarray}
L_{gf}  & = &
\biggl( b^\mu \omega_{\mu \nu} F^\nu
+ \pi^\mu \zeta_{\mu \nu } f^\nu
+ d^\mu \varsigma _{\mu \nu }\rho^\nu
+ \lambda^{\mu \nu} \varrho_{\mu \nu \alpha \beta} f^{\alpha \beta}
\nonumber \\
&&
- \frac{1}{2} b^\mu \omega_{\mu \nu } b^\nu
- \frac{1}{2} \pi^\mu \zeta_{\mu \nu } \pi^\nu
- \frac{1}{2} d^\mu \varsigma_{\mu \nu } d^\nu
- \frac{1}{2} \lambda^{\mu \nu} \varrho_{\mu \nu \alpha \beta}
\lambda^{\alpha \beta} \biggr) e
\label{tetrad}
\end{eqnarray}
where  $\{ b_\mu$, $\pi_\mu, d_\mu, \lambda_{\mu \nu} \}$
are auxiliary fields,
$\{ \omega_{\mu \nu }, \varsigma_{\mu \nu }, \zeta_{\mu \nu }
\varrho_{\mu \nu \alpha \beta} \}$ are arbitrary operators and
$\{ F_\mu$, $f_{\mu \nu }$, $f_\mu$, $\rho_\mu \}$ fix the general coordinate,
local Lorentz, projective and antisymmetric gauges. Since auxiliary
fields appear without derivatives in the Lagrangian, they can be
eliminated by means of their equations of motion which yield

\begin{equation}
L_{gf}  =
\biggl( \frac{1}{2} F^\mu \omega_{\mu \nu } F^\nu
+ \frac{1}{2} f^\mu \zeta_{\mu \nu } f^\nu
+ \frac{1}{2} \rho^\mu \varsigma_{\mu \nu } \rho^\nu
+ \frac{1}{2} f^{\mu \nu} \varrho_{\mu \nu \alpha \beta}
f^{\alpha \beta} \biggr) e
\end{equation}
We fix the coordinate, projective, antisymmetric and local Lorentz
gauges by means of the following terms:

\begin{equation}
F^\mu = e^{-\tau} \partial_\nu \left( e_a^{~\mu} e^{a \nu} e^\tau \right)
+ a_1 \bar{\Omega}^{\mu \alpha}_{~~\alpha}
+ a_2 \bar{\Omega}^{a \mu}_{~~a}
+ a_3 \bar{\Omega}^{a~\mu}_{~a}
\end{equation}

\begin{equation}
f^\lambda = \left(f_1 e_a^\lambda e^{b \nu}
+  f_2 e^{b \lambda} e^\nu_a
+ f_3 g^{\nu \lambda} \delta^a_b \right)
\bar{\Omega}^a_{~b \nu}
\end{equation}
where $\tau$ and $a_1,a_2, a_3$ are an arbitrary constants and
$f_i$ are the constants satisfying the condition (\ref{aux}),

\begin{equation}
\rho^\lambda = \epsilon^{\lambda \mu \sigma \nu} \bar{\Omega}_{a b \nu}
e^a_\mu e^b_\sigma
\end{equation}

\begin{eqnarray}
f_{ab} & = & c_1 \partial_\mu \left( \bar{\Omega}_{a~ b}^{~\mu}
-  \bar{\Omega}_{b ~a}^{~\mu} \right)
+ c_2 \partial_\mu \left( \bar{\Omega}_{a b}^{~~\mu}
-  \bar{\Omega}_{b a}^{~~\mu} \right)
\nonumber \\
& + &
c_3 \partial_\mu \left( \bar{\Omega}_{~a b}^{\mu}
-  \bar{\Omega}_{~b a}^{\mu} \right)
+ c_4 \left( e_{ab} - e_{ba} \right)
\end{eqnarray}
where $\{ Ó_i \}$ are arbitrary constants.

The most general coordinate and projective gauge fixing terms are

\begin{equation}
F^\mu = e^{-\tau} \partial_\nu \left( e_a^{~\mu} e^{a \nu} e^\tau \right)
+ a_1 \bar{\Omega}^{\mu \alpha}_{~~\alpha}
+ a_2 \bar{\Omega}^{a \mu}_{~~a}
+ a_3 \bar{\Omega}^{a~\mu}_{~a}
+ M^{\mu \alpha \beta ~ b \nu}_{~~~~b}
\nabla_\alpha \nabla_\beta \bar{\Omega}^a_{~b \nu}
\end{equation}

\begin{equation}
f^\lambda = \left(f_1 e_a^\lambda e^{b \nu}
+  f_2 e^{b \lambda} e^\nu_a
+ f_3 g^{\nu \lambda} \delta^a_b \right)
\bar{\Omega}^a_{~b \nu}
+ N^{\mu \alpha \beta ~ b \nu}_{~~~~b}
\nabla_\alpha \nabla_\beta \bar{\Omega}^a_{~b \nu}
\end{equation}
where $M^{\mu \rho \lambda ~ \alpha \beta}_{~~~~\sigma} $ and
$N^{\mu \rho \lambda ~ \alpha \beta}_{~~~~\sigma} $ are
a product of metric tensors and Kroneker's symbols.
To avoid the problems with unitarity of the model, we consider the case
$M^{\mu \rho \lambda ~ \alpha \beta}_{~~~~\sigma} =
N^{\mu \rho \lambda ~ \alpha \beta}_{~~~~\sigma}  = 0 $.

The complete BRST-transformations are

\begin{eqnarray}
{\bf s} e^a_{~ \mu} & = & {\it k} \left(
e^a_{~\sigma} \partial_\mu c^\sigma +
c^\sigma \partial_\sigma e^a_{~\mu} + \Theta^a_{~b} e^b_{~\mu} \right)
+ O ({\it k}^2)
\nonumber \\
{\bf s} \bar{\Omega}^a_{~b \mu}  & = & {\it k} \left(
\bar{\Omega}^a_{~b \sigma} \partial_\mu c^\sigma
+ c^\sigma \partial_\sigma \bar{\Omega}^a_{~b \mu}
+ \bar{\nabla}_\mu \Theta^a_{~b}
+ \delta^a_b  \chi_\mu + \eta^a_{~ b m } \right) + O ({\it k}^2)
\nonumber \\
{\bf s} \bar{c}_\mu  & = & b_\mu
~~~~~~~~~~{\bf s} b_\mu  =  0
~~~~~~~{\bf s} c^\mu   =  c^\lambda \partial_\lambda c^\mu
\nonumber \\
{\bf s} \bar{\chi}_\mu & = & \pi_\mu
~~~~~~~~~~{\bf s} \pi_\mu = 0
~~~~~~{\bf s} \chi^\nu = 0
\nonumber \\
{\bf s} \bar{\eta}_\mu & = & d_\mu
~~~~~~~~~~{\bf s} d_\mu = 0
~~~~~~{\bf s} \eta^\nu = 0
\nonumber \\
{\bf s} \bar{\Theta}_a^{~b} & = & \lambda_a^{~b}
~~~~~~~~{\bf s} \lambda_a^{~b} = 0
~~~~~~{\bf s} \Theta^a_{~b} = c^\sigma \partial_\sigma \Theta^a_{~b}
+ \Theta^a_{~m} \Theta^m_{~b}
\label{tetrad-BRST}
\end{eqnarray}
where
${\bf s}$ is a graded, nilpotent BRST operator and
$\{\bar{c}_\nu, c^\mu,\}$, $\{ \bar{\chi}_\alpha, \chi^\beta \}$,
$\{\bar{\eta}_\sigma, \eta^\lambda_{~\mu \nu } \}$,
$ \{\bar{\Theta}_m^{~n}, \Theta^a_{~b} \} $
are anticommuting ghost fields connected with the general coordinate,
projective, antisymmetric and the local Lorentz transformations,
respectively.

The quantum Lagrangian is
\begin{eqnarray}
L_{quant} = L_{clas}
+ {\bf s} \Biggl( \bar{c}^\mu \omega_{\mu \nu}
 \left( F^\nu - \frac{1}{2} b^\nu \right)
 + \bar{\chi}^\mu \zeta_{\mu \nu}
 \left( f^\nu - \frac{1}{2} \pi^\nu \right)
 \nonumber \\
  +
 \bar{\eta}^\mu \varsigma_{\mu \nu}
 \left( \rho^\nu - \frac{1}{2} d^\nu \right)
 + \bar{\Theta}^{\alpha \beta} \varrho_{\alpha \beta \mu \nu}
 \left( f^{\mu \nu } - \frac{1}{2} \lambda^{\mu \nu} \right)
 \Biggr)
\end{eqnarray}
where for simplicity we consider the case ${\bf s} \omega_{\mu \nu } =
{\bf s} \zeta_{\mu \nu } = {\bf s} \varsigma_{\mu \nu } =
{\bf s} \varrho_{\alpha \beta \mu \nu } =0$. The generating functional has
the following form

\begin{eqnarray}
e^{i W} & = &
\int de^a_{~\mu }~ d\bar{\Omega}^a_{~b \nu }
~d \bar{\Theta}^{ab} ~d \Theta^{mn}
~d \bar{c}^\mu ~dc^\nu ~d \bar{\chi}^\mu ~d\chi^\nu
~d \bar{\eta}^\mu ~d\eta^{\sigma \lambda \nu}
e^{i S_{quan}}
\nonumber \\
&&
\left(det \omega_{\mu \nu } \right)^{\frac{1}{2}}
\left(det \zeta_{\mu \nu } \right)^{\frac{1}{2}}
\left(det \varsigma_{\mu \nu } \right)^{\frac{1}{2}}
\left(det \varrho_{abmn} \right)^{\frac{1}{2}}
\end{eqnarray}

The ghost Lagrangian is

\begin{eqnarray}
 L_{gh}  & = &
- \bar{c}^\mu \omega_{\mu \nu} {\bf s} F^\nu
- \bar{\chi}^\mu \zeta_{\mu \nu } {\bf s} f^\nu
- \bar{\eta}^\mu \varsigma_{\mu \nu } {\bf s} \rho^\nu
- \bar{\Theta}^{\mu \nu} \varrho_{\mu \nu  \alpha \beta} {\bf s}
f^{\alpha \beta} =
\nonumber \\
& - &
\left( \bar{c}^\mu ~\bar{\chi}^\mu ~\bar{\eta}^\mu
~\bar{\Theta}^{ab}\right)
\left( \begin{array}{cccc}
\omega_{\mu \alpha} \triangle^\alpha_{~\nu} & B \omega_{\mu \nu}
& 0 & K\omega_{\mu m} \bar{\nabla}_n \\
\zeta_{\mu \alpha} Z^\alpha_{~\nu} &  A \zeta_{\mu \nu } & 0 &
L\zeta_{\mu m} \bar{\nabla}_n \\
\varsigma_{\mu \alpha}D^\alpha_{~\nu} & 0 &
\varsigma_{\mu \alpha} \epsilon^{\alpha \sigma \lambda \nu} &
\varsigma_{\mu \alpha} \epsilon^{\alpha \beta m n} \bar{\nabla}_\beta \\
\varrho_{abij} M^{ij}_{~~\nu} & U \varrho_{ab \mu \nu} \partial_\mu
& T \varrho_{ab \lambda \nu} \partial_\sigma & \varrho_{abij}S^{ij}_{~~mn}
\end{array} \right)
\left( \begin{array}{c}
c^\nu \\
\chi^\nu \\
\eta_{\sigma \lambda \nu } \\
{\Theta^{mn}}
\end{array} \right)
\nonumber \\
\label{full-ghost}
\end{eqnarray}
where  $A = f_1 + f_2 + 4f_3$, $B = a_1 + a_2 + 4 a_3$,
$K = (a_1-a_2) $, $L = (f_1 - f_2 ) $, $T = 2 (c_2 + c_3 - c_1 )$,
$U = 2(c_1 + c_3)$,
and $\triangle^\mu_{~\nu}, D^\mu_{~ \nu}, Z^\mu_{~\nu}, M^a_{~b \nu}$
and $S^{ab}_{~~mn}$ are

\begin{eqnarray}
\triangle_{\mu \nu} & = &
{\it k}  \biggl( - g_{\mu \nu} \nabla^2
+ \frac{1}{2} \left( \tau - 1 \right)
\left( \nabla_\mu \nabla_\nu + \nabla_\nu \nabla_\mu \right)
- \frac{1}{2} \left( 1 + \tau \right) R_{\mu \nu}
\nonumber \\
&&
+ a_1 \left( \bar{\Omega}^{~\sigma }_{\mu ~\nu} \partial_\sigma
+ \partial_\nu \bar{\Omega}_{\mu \sigma}^{~~~ \sigma} \right)
+ a_2 \left( \bar{\Omega}^\sigma_{~ \mu \nu} \partial_\sigma
+ \partial_\nu \bar{\Omega}^\sigma_{~ \mu \sigma} \right)
\nonumber \\
&&
+ a_3 \left( \bar{\Omega}^\sigma_{~ \sigma \nu} \partial_\mu
+ \partial_\nu \bar{\Omega}^\sigma_{~ \sigma \mu} \right)
\biggr) + O({\it k}^2)
\end{eqnarray}

\begin{eqnarray}
Z_{\mu \nu} & = & {\it k} \biggl(
f_1 \left( \bar{\Omega}^{~ \sigma}_{\mu ~ \nu} \partial_\sigma
+ \partial_\nu \bar{\Omega}_{\mu \sigma}^{~~ \sigma} \right)
+ f_2 \left( \bar{\Omega}^\sigma_{~ \mu \nu} \partial_\sigma
+ \partial_\nu \bar{\Omega}^\sigma_{~ \mu \sigma} \right)
\nonumber \\
&&
+ f_3 \left( \bar{\Omega}^\sigma_{~ \sigma \nu} \partial_\mu
+ \partial_\nu \bar{\Omega}^\sigma_{~\sigma \mu} \right)
\biggr) + O({\it k}^2)
\end{eqnarray}

\begin{equation}
D^\mu_{~ \nu}  =  {\it k} \epsilon^{\mu \alpha \beta \lambda}
\left( \bar{\Omega}_{\alpha \beta \nu} \partial_\lambda
+ \partial_\nu \bar{\Omega}_{\alpha \beta \lambda}
\right) + O({\it k}^2)
\end{equation}

\begin{eqnarray}
S_{abmn} & = &
\left( \eta_{am} \eta_{bn} - \eta_{an} \eta_{bm} \right)
\left( c_2 \partial_\mu \bar{\nabla}_\mu  + c_4 \right)
\nonumber \\
& + & \frac{\left( c_1 - c_3 \right)}{2}
\left( \partial_n \bar{\nabla}_b \eta_{m a}
- \partial_m \bar{\nabla}_b \eta_{n a}
- \partial_n \bar{\nabla}_a \eta_{m b}
+ \partial_m \bar{\nabla}_a \eta_{b n}
\right)
\end{eqnarray}

\begin{eqnarray}
M_{ab \nu} & = &
c_1 \left( \partial_\mu \left(
\bar{\Omega}^{~\mu}_{a ~ \nu} \partial_b
- \bar{\Omega}^{~\mu }_{b ~ \nu} \partial_a
+ \partial_\nu \bar{\Omega}^{~\mu }_{a ~ b}
- \partial_\nu \bar{\Omega}^{~\mu}_{b ~ a} \right) \right)
\nonumber \\
& + &
c_2 \left( \partial_\mu \left(
\bar{\Omega}_{a b \nu} \partial_\mu
- \bar{\Omega}_{b a \nu} \partial_\mu
+ \partial_\nu \bar{\Omega}_{a b}^{~~~ \mu}
- \partial_\nu \bar{\Omega}_{b a}^{~~~ \mu} \right) \right)
\nonumber \\
& + &
c_4 \left(
\eta_{a \nu} \partial_b
- \eta_{\nu a } \partial_a
+ \partial_\nu e_{a b}
- \partial_\nu e_{b a} \right) + O({\it k}^2)
\end{eqnarray}

Let us consider the case $B = K = L = T = U = 0$. Define  new variables

\begin{eqnarray}
\tilde {\chi}^\nu & = & \chi^\nu + \frac{1}{A} Z^\nu_{~\sigma} c^\sigma
\nonumber \\
\tilde {\eta}^{\sigma \lambda \nu } & = & \eta^{\sigma \lambda \nu }
+ \frac{1}{6} \epsilon^{\sigma \lambda \nu \mu }Z_{\mu \alpha } c^\alpha
\nonumber \\
\tilde {\Theta}^a_{~b} & = & \Theta^a_{~b} + M^a_{~b \sigma} c^\sigma
\end{eqnarray}
This redefinition does not change the functional integral measure.
In the new variables the ghost Lagrangian has the diagonal form:

\begin{equation}
 L_{gh} =
\left( \bar{c}^\mu ~\bar{\chi}^\mu ~\bar{\eta}^\mu
~\bar{\Theta}^{ab}\right)
\left( \begin{array}{cccc}
\omega_{\mu \alpha} \triangle^\alpha_{~\nu} & 0 & 0 & 0 \\
0 &  A \zeta_{\mu \nu } & 0 & 0 \\
0 & 0 & \varsigma_{\mu \alpha} \epsilon^{\alpha \sigma \lambda \nu} & 0 \\
0 & 0 & 0 & \varrho_{abij}S^{ij}_{~~mn}
\end{array} \right)
\left( \begin{array}{c}
c^\nu \\
\tilde {\chi}^\nu  \\
\tilde {\eta}_{\sigma \lambda \nu }  \\
\tilde {\Theta}^{mn}
\end{array} \right)
\label{res}
\end{equation}

We see that in the tetrad formalism the projective and antisymmetric
ghosts also give the contribution in the effective action. The
one-loop generating functional is

\begin{eqnarray}
e^{i W} & = &
\int de^a_{~\mu }~ d\bar{\Omega}^a_{~b \nu }
~d \bar{\Theta}^{ab} ~d \Theta^{mn}
~d \bar{c}^\mu ~dc^\nu ~d \bar{\chi}^\mu ~d\chi^\nu
~d \bar{\eta}^\mu ~d\eta^{\sigma \lambda \nu}
e^{i \left( S_{clas} + S_{gf} \right) }
\nonumber \\
&&
\left(det \omega_{\mu \nu } \right)^{\frac{1}{2}}
\left(det \zeta_{\mu \nu } \right)^{\frac{3}{2}}
\left(det \varsigma_{\mu \nu } \right)^{\frac{3}{2}}
\left(det \varrho_{abmn} \right)^{\frac{1}{2}}
\left(det \omega_{\mu \alpha} \triangle^\alpha_{~\nu} \right)
\left(det \varrho_{abij}S^{ij}_{~~mn} \right)
\label{result2}
\end{eqnarray}
This contribution is added to the corresponding Nielsen-Kallosh ghost one.

Consider the question of anomalies, related to transformations
(\ref{testtr}).  Anomaly is violation of some classical symmetries at
the quantum level.  Anomalies (like the well known Adler-Bell-Jackiw
anomaly) may arise in the case when the variation of the action
explicitly depends on the space-time dimension. If there is no such
dependence, the dimensional regularization retain the symmetry of the
regularized model. The action principles \cite{action}, guarantees that
the minimal subtraction scheme \cite{min} will not break the symmetry
after renormalization.

Transformations (\ref{testtr}) and the variation of the affine-metric  action
do not explicitly depend on the space-time dimension. Consequently, there is
no anomaly in the theory associated with these transformations.

\section{Conclusion}

A lot of unsolved problems in quantum gravity make one to search for new
ways to solve them. At present, no model is quite satisfactory from the
viewpoint of quantum field theory, possessing unitarity, renormalizability,
existing $S$-matrix, etc. A criterion of the physical significance of the
results of loop calculation can be the gauge and parametric
independence \cite{MKL2}.

In the present paper, we have considered the affine-metric gravity with
extra local symmetries (\ref{testtr}) related to the transformations
of affine connection.  These symmetries do not have the "ordinary"
physical meaning and geometrical nature: for the extra local symmetries
no additional gauge fields appear. The role of these symmetries is to
suppress the counterterms that break the renormalizability of the model
and restrict arbitrariness of the initial Lagrangian. Besides, these
new symmetries imposes new constraints on the source term.
Although they are local symmetries we have shown that no
corresponding anomalies are generated. We have constructed the
BRST-transformations connected with extra symmetries in geometric
(\ref{full-BRST}) and tetrad (\ref{tetrad-BRST}) approaches and shown
that in both formalisms these symmetries give the additional contribution
to the effective action (see (\ref{result1}) and
(\ref{result2}), respectively) which is proportional to the
corresponding Nielsen-Kallosh ghost one.  These additional
contributions may improve the renormalization properties of the
theory.

One of the unresolved problems is to find the full set of extra
symmetries, existing in affine-metric gravity.
We know only two ways to solve this problem. The first one is to
postulate that these extra symmetries have some physical sense
\footnote {In fact, these extra symmetries give rise to some
deformation of the initial connection. This deformed connection can
be used for construction of the Grand Unified Theory in the framework
of the supergravity.}.  The other way is to finds the full set of
first-class constraints using the hamiltonian formalism
\cite{Bars}. But this method does not allow us to
understand the physical ground of extra local symmetries.

We don't know the full set of extra local symmetries in affine-metric
gravity. In present paper we discussed only two kinds of these
symmetries :  the projective (\ref{projec}) and antisymmetric
(\ref{ant}) ones. There exist two interesting particular cases of the
considered extra symmetries. In these cases the parameter of extra
transformation is the derivative of some field \cite{Nev1},
\cite{Bars}:

\begin{equation}
\bar{\Gamma}^\sigma_{~\mu \nu}  \rightarrow
'\bar{\Gamma}^\sigma_{~\mu \nu} = \bar{\Gamma}^\sigma_{~ \mu \nu} +
\delta^\sigma_\mu \partial_\nu \Phi(x)
+ g^{\sigma \alpha } \partial_{[ \alpha} \lambda_{\mu \nu ]}(x)
\end{equation}
The general result of this paper does not change for these cases:
it is necessary to fix the extra local symmetries and then these
symmetries will give the non-trivial contribution to the one-loop
divergences. To proof this statement one needs only to modify a
little the gauge conditions (\ref{general}) and (\ref{axial}).

\begin{center}
{\bf Acknowledgments}
\end{center}
I would like to thank L. Avdeev for help in editing the text.  I would
like to express my gratitude to D. I. Kazakov and I. L. Buchbinder for
valuable discussions. I am indebted to G. Sandukovskaya and
A. Gladyshev for critical reading of the manuscript. I am very
grateful to L.  Averchenkova for moral support when the paper was
written.

\end{document}